\newcommand{\ba}{\begin{array}}
\newcommand{\ea}{\end{array}}
\newcommand{\bean}{\begin{eqnarray*}}
\newcommand{\eean}{\end{eqnarray*}}
\newcommand{\rm}[1]{\mathrm{#1}}

\newcommand{\ep}{\epsilon}
\newcommand{\p}{\partial}

\newcommand{\Tr}{\mathrm{Tr}}

\newcommand{\bra}[1]{\left\langle {#1} \right|}
\newcommand{\ket}[1]{\left|  #1 \right\rangle}

\newcommand{\bmx}[1]{\left(\begin{array}{*{#1}{c}}}
\newcommand{\emx}{\end{array}\right)}
\newcommand{\bmxw}[1]{\renewcommand{\arraystretch}{2}\left(\begin{array}{*{#1}{c}}}
\newcommand{\bmxww}[1]{\renewcommand{\arraystretch}{2.5}\left(\begin{array}{*{#1}{c}}}
\newcommand{\bdet}[1]{\renewcommand{\arraystretch}{1.2}
	\left|\begin{array}{*{#1}{c}}}
\newcommand{\edet}{\end{array}\right|\renewcommand{\arraystretch}{1}}
\newcommand{\beq}{\begin{equation}}
\newcommand{\eeq}{\end{equation}}
\newcommand{\bea}{\begin{eqnarray}}
\newcommand{\eea}{\end{eqnarray}}

\newcommand{\ditem}[1]{\item[$\diamond$]}

\newcommand{\bit}{\begin{itemize}}
\newcommand{\eit}{\end{itemize}}

\newcommand{\eab}{\begin{eqnarray}}
\newcommand{\eae}{\end{eqnarray}}

\documentclass[aps,prb,twocolumn,superscriptaddress]{revtex4-1}
\usepackage{graphics}
\usepackage{epsfig}
\usepackage{amssymb}
\usepackage{color} 
\usepackage{subfigure}
\usepackage{ulem}

\begin{document}


\title{When does Magnetization Depend on the Boundary?}


\author{Kuang-Ting Chen}
\affiliation{Department of Physics, Massachusetts Institute of Technology, Cambridge, MA 02139}
\author{Patrick A. Lee}
\affiliation{Department of Physics, Massachusetts Institute of Technology, Cambridge, MA 02139}


\date{\today}

\begin{abstract}
In this paper, we investigate in general how thermodynamic quantities such as the polarization, magnetization and the magneto-electric tensor are affected by the boundaries. We show that when the calculation with periodic boundary conditions does not involve a Berry's phase, the quantity in question is determined unambiguously by the bulk, even in the presence of gapless surface states. When the calculation involves a Berry's phase, the bulk can only determine the quantity up to some quantized value, given that (i) there are no gapless surface states, (ii) the surfaces do not break the symmetries preserved by the bulk, and (iii) the system is kept at charge neutrality. If any of the above conditions is violated, the quantity is then determined entirely by the details at the boundary. For components of response functions such as the magneto-electric tensor that are entirely determined by the bulk, the value of those components should connect smoothly to finite frequency and momentum.

\end{abstract}


\maketitle


\section{Introduction}
Recently there has been a resurging interest in understanding the general orbital magneto-electric (ME) response\citep{Essin2009,Essin2010,Malashevich2010,Coh2011}. This is due to the fact that the isotropic magneto-electric effect or the so-called $\theta$-term, $\mathcal{L}_\theta=(\theta e^2/2\pi h)E\cdot B$ with $\theta=\pi$, is suggested to describe the the time reversal invarint (TRI) topological band insulator (TBI) in three spatial dimensions\cite{Qi08} (3d). Usually the signature of the 3d TBI is the existence of an odd number of Dirac cones on the boundary surfaces, when the time reversal symmetry (TRS) is preserved.\citep{Hasan2010} When the surface states are gapped out by breaking the TRS locally on the boundaries, a half integer Quantum Hall effect will take place, and give rise to a quantized bulk magneto-electric response.\cite{Qi08} It is later shown that this isotropic response is only a part of the more general anisotropic orbital ME tensor defined in the bulk\cite{Essin2010}:
\beq
\alpha_{ij}=\alpha_{\rm \theta}\delta_{ij}+\alpha_{{\rm 3d}ij},
\eeq
we define $\alpha_\theta=\frac13\Tr\alpha_{ij}$ and $\alpha_{\rm 3d}$ is therefore traceless. $\alpha_{ij}$ describes either the orbital magneto-polarizability (OMP) or the orbital electric susceptibility (OES):
\beq
 \alpha_{ij}={\rm d}P_i/{\rm d}B_j={\rm d}M_j/{\rm d}E_i;
\eeq
OMP and OES are equal via a Maxwell relation. 
 
One peculiarity of the ME tensor is that $\alpha_\theta$ is only determined up to integer multiples of $e^2/h$ by the bulk band structure.\cite{Malashevich2010,Essin2010} The specific value of $\alpha_\theta$ depends on the details at the boundary. From the polarization perspective, if we attach an integer quantum Hall (IQH) layer with filling $\nu=\pm 1$ respectively on the top and bottom surface of a cylinder, the orbital magneto-polarizability (OMP), i.e., ${\rm d}P/{\rm d}B$, along the axis of the cylinder will change by $e^2/h$, due to the density locking to the magnetic field of the top and the bottom IQH layer. From the magnetization perspective, if we attach an IQH layer with filling $\nu=1$ on the side surfaces of the cylinder, the orbital electric susceptibility (OES), ${\rm d}M/{\rm d}E$, will also change by $e^2/h$, due to the extra Hall current flowing on the surface in response to the electric field. Either way, the ME effect is only determined up to an integer multiple of $e^2/h$. In the bulk, this ambiguity corresponds well to the fact that $\theta$ as a coefficient in front of $(E\cdot B)$ is an angle only defined up to integer multiple of $2\pi$, because $\int E\cdot B {\rm d}^3 x{\rm d}t$ is quantized. $\alpha$ is odd under TRS, but this ambiguity makes it possible for $\alpha_\theta$ not to vanish with TRS preserved, as $\alpha_\theta=\pm e^2/2h$ is differed from each other by $e^2/h$. $\alpha_{\theta}=0$ and $\alpha_{\theta}=e^2/2h$ then describe two different insulating states of matter under TRS.

 However, the above properties raise some questions. From the polarization perspective this ambiguity from the bulk is not so surprising, since the zero field polarization is already ambiguous with periodic boundary conditions.\cite{King93,Zak1989} From the magnetization perspective, however, this ambiguity is a bit more puzzling, because one commonly regards magnetization as a bulk property.\cite{Thonhauser05,Ceresoli06} In particular, with periodic boundary conditions there seems to be no reason to expect any ambiguity in the magnetization, whereas the ambiguity in the polarization is easily understood. 

Before answering this rather specific question, we note that there is a even more general one: to what extent are thermodynamic quantities such as polarization, magnetization, and ME response determined by the bulk? Unlike the conventional thermodynamic quantities which are entirely independent of the boundary, we have already seen that boundary sometimes plays a role. How can we tell when will a thermodynamic variable depend on the boundary and when will it not?

In the following, we will discuss case by case from the ground state polarization, orbital magnetization, to the magneto-electric tensor. We will argue through Gedanken experiments that some of them depends on the boundary while others don't. We will verify our argument with numerical simulations. By matching the observations with our previous calculation done with periodic boundary conditions, we can then directly tell from the calculation with periodic boundary conditions how different thermodynamic quantities depend on boundaries.

\section{Ground state polarization}
The ground state polarization is given by the following formula with periodic boundary conditions:\cite{King93}
\beq
P=-ie\int_{BZ}\frac{{\rm d}^d k}{2\pi^d}\sum_{\alpha(k)\in occ}\bra{\alpha(k)}\frac{\p}{\p k}\ket{\alpha(k)}.
\eeq
In one spatial dimension (1d), the polarization is defined modulo $e$ with periodic boundary conditions: $P=P_0+ne$, with $n$ an integer. This corresponds to the observation that with periodic boundary conditions, we can move every electron to the next unit cell and return to the original state, while the two states should by definition have polarization differed by $e$. With two ends, the polarization will take one specific value, depending on the number of charges we put at the two ends.

However, if there are zero modes at the two ends, the polarization is then ambiguous, as theoretically we can consider superposition of states of different occupancy of the zero modes. The bulk value of the polarization thus depends entirely on the boundary.

In 3d it is a bit more interesting. For simplicity let us assume the system sits on a cubic lattice of size $a$. Now the bulk formula has an ambiguity of $e/a^2$, which also corresponds well to the fact that we can move every electron to the next unit cell and return to the same state. However, with boundary surfaces the situation becomes quite different. Consider a capacitor setup. We are allowed to put any number of charges on each of the opposing surfaces, resulting in a change of the polarization in units of $e/A$ (A is the total surface area). In the thermodynamic limit, we can put any finite density of charges on the surface, and the polarization in the bulk can take any value. Our bulk formula is thus no longer valid. To accommodate the charge on the surface, however, the system needs to either be in a metallic state near the boundary, or to break the lattice translation symmetry in the two in-plane directions. If neither condition is satisfied, then we can only add an integer number of electrons per unit cell, and the bulk formula is recovered, with the remaining ambiguity determined by the surface. 

How can the bulk formula become invalid? We note that the ground state polarization can be understood as a Berry's phase when one adiabatically turns on the electric field. Firstly, in order for the Berry's phase to make any sense, the system has to be gapped. This is the reason why a metallic surface can render the bulk formula invalid. Secondly, if we break the lattice translation symmetry in the two directions perpendicular to the electric field, we can no longer integrate over the momentum in those directions but should instead sum over a large number of sub-bands labelled by the remaining momentum along the direction of the electric field. The polarization will have an ambiguity of $e/A$ in this case. This is different from the conventional thermodynamic quantity, which will require a symmetry breaking in the bulk to change its value. The Berry's phase is thus a rather fragile thermodynamic quantity.

\section{Ground state orbital magnetization}
It is not immediately obvious that the orbital magnetization is independent of the boundary. In the bulk the operator $\hat M\propto({\bf r\times v})$ is ill-defined with periodic boundary conditions, and seems to be growing as one goes near the boundary. Indeed, when one numerically compute $\langle\hat M\rangle$ summing over the local orbitals, there is a finite contribution from the boundary orbitals, which renders the total orbital magnetization different from the naive bulk value.~\cite{Ceresoli06} Nevertheless, it has been shown\cite{Ceresoli06} that the boundary contribution is in fact independent of the details at the boundary via the use of local Wannier functions, in an insulator with zero Chern number. 

However, in a Chern insulator, a local Wannier function can not be found\cite{Thonhauser06,Thouless1984}, because the Bloch functions cannot be periodic and smoothly defined over the Brillouin Zone. To see that even in this case the orbital magnetization is still independent of the boundaries, we can consider the following setup:

Suppose we have an insulator with a non-vanishing Chern number in two dimensions. Let us imagine putting an auxiliary layer of insulator on top, with an opposite Chern number, without any interaction with the original one. The new insulator as a whole is then of total Chern number zero. We can therefore make a local Wannier orbital, by a linear combination of orbitals from the two layers.\cite{Soluyanov2011} The argument then goes through for the insulator as a whole, and the total orbital magnetization should be independent of the boundary. Now since there is no interaction between the two layers, the total magnetization is just the sum of the magnetization of the original insulator and the auxiliary insulator. We now consider a particular boundary condition, where the two insulators couples to independent boundary terms that do not interact with each other as well. Let us only vary the boundary terms that couple to the original insulator. The total magnetization cannot change, and neither the contribution from the auxiliary insulator. We thus have to conclude that even for a Chern insulator, the orbital magnetization is independent of the boundaries. 

From this abstract point of view, the generalization to Chern insulators seems rather trivial. However, the presence of gapless chiral edge states may cause one to worry. Suppose we can gate the material to supply a constant chemical potential, what will happen if we turn up the electric potential on the edge? Will the edge current decrease because fewer edge states are occupied, or will it stays the same as required for the bulk magnetization not to change?

\begin{figure}[htb]
	\centering
	\subfigure[]{\includegraphics[width=3.5cm]{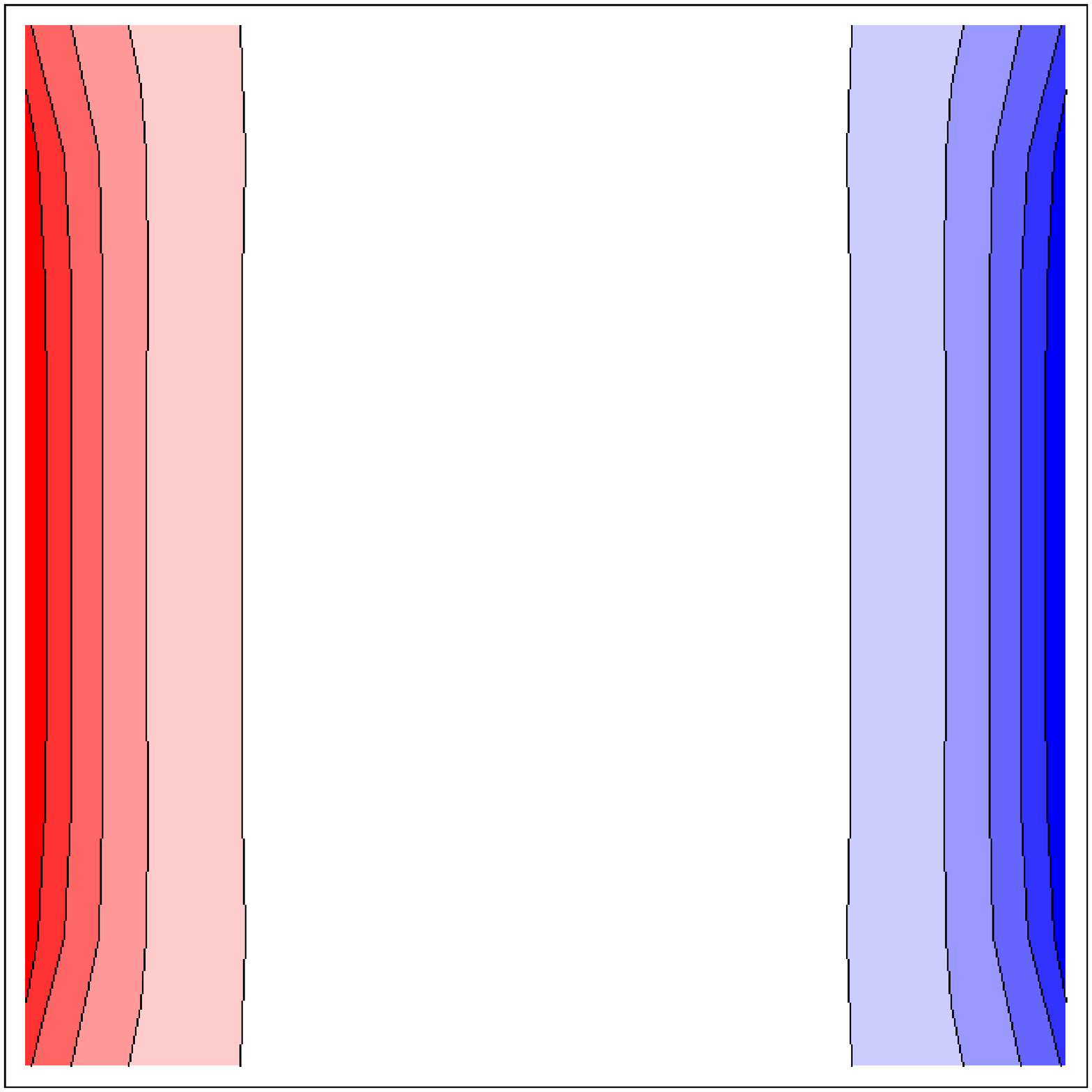}}
	\subfigure[]{\includegraphics[width=3.5cm]{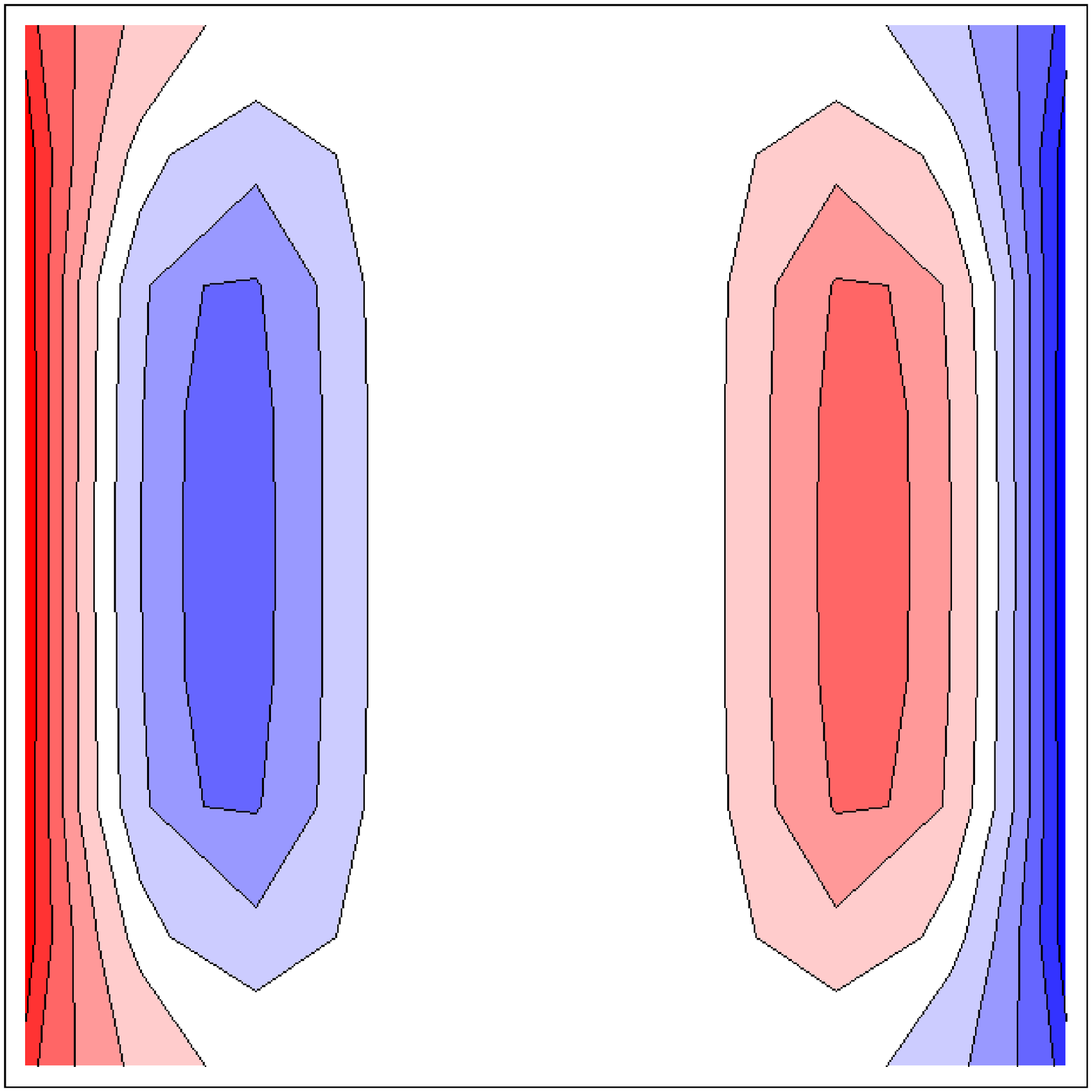}}
   \subfigure{\includegraphics[width=1cm]{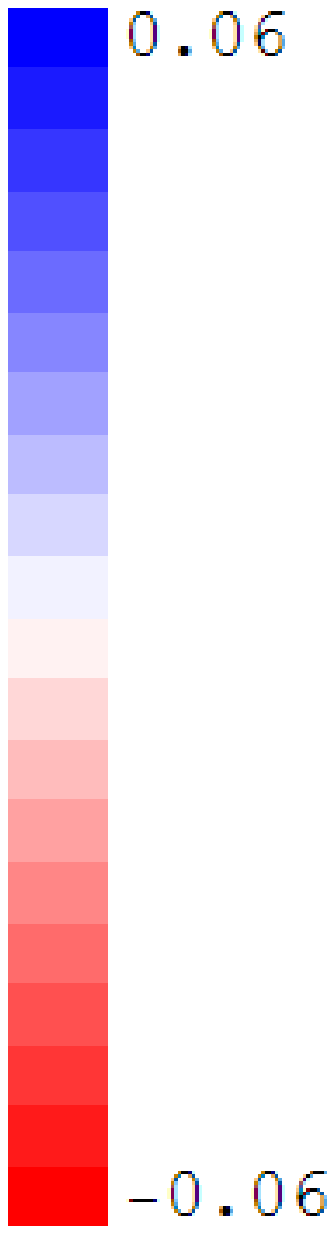}}
	\caption{We take our Hamiltonian to be \\$H=\sum_n c_n^\dag(\tau_z-i\tau_x)c_{n+\hat x}+c_n^\dag(\tau_z-i\tau_y)c_{n+\hat y}+mc_n^\dag\tau_zc_n+h.c.$,\\ where $\tau's$ are the Pauli matrices. At half filling with $m=1.5$, the band carries a Chern number $C_1=1$. If we set the chemical potential $\mu=0$, the ground state has no magnetization. We put the Hamiltonian on a $10\times 10$ lattice, and take open boundary conditions in both directions. The current on the vertical links is plotted. We relate the current to the magnetization by $I^b=\ep^{ab}\p_aM$, and take the magnetization at the middle to represent the bulk magnetization. (a) $\mu=0.5$. As expected, some edge states are occupied and give rise to a bulk magnetization. (b) If we set $\mu=0$ but locally apply an electric potential $V=-0.5$ to the first two rows at the boundary, the edge states are again occupied. However, in the region next to those layers, a counterpropagating current takes place. The bulk magnetization remains zero (barring some finite size effect).
}
	\label{chern}
\end{figure}  

We do a straightforward numerical simulation to resolve this paradox. The result is shown in Fig. \ref{chern}. We can see that while shifting the overall chemical potential creates circulating currents, altering the electric potential locally at the edge does not change the bulk magnetization. If we look closer, while the current right at the edge is changed, there is a counter-propagating current near the edge, which keeps the total current localized near one edge constant. The counter-propagating current is just the integer quantum Hall response to the electric potential gradient. This bulk quantum Hall current exactly compensates for the current carried by the now-unoccupied edge states, and leaves the bulk magnetization insensitive to the change of the potential local near the edge.


A very similar puzzle arises in the $S_z$ conserved spin Hall insulator. On the edge there are counter-propagating TR-paired edge states. When we apply a uniform Zeeman field $H_z$, there will be a net circulating current from the edge states. We can therefore deduce a bulk orbital magnetization response to the Zeeman field. We call this the orbital-Zeeman susceptibility. However, one can locally break the $S_z$ conservation together with the TR symmetry near the edge, to gap out the edge states. In this case, will there still be a bulk magnetization response to the Zeeman field?

\begin{figure}[htb]
	\centering
	\subfigure[]{\includegraphics[width=4cm]{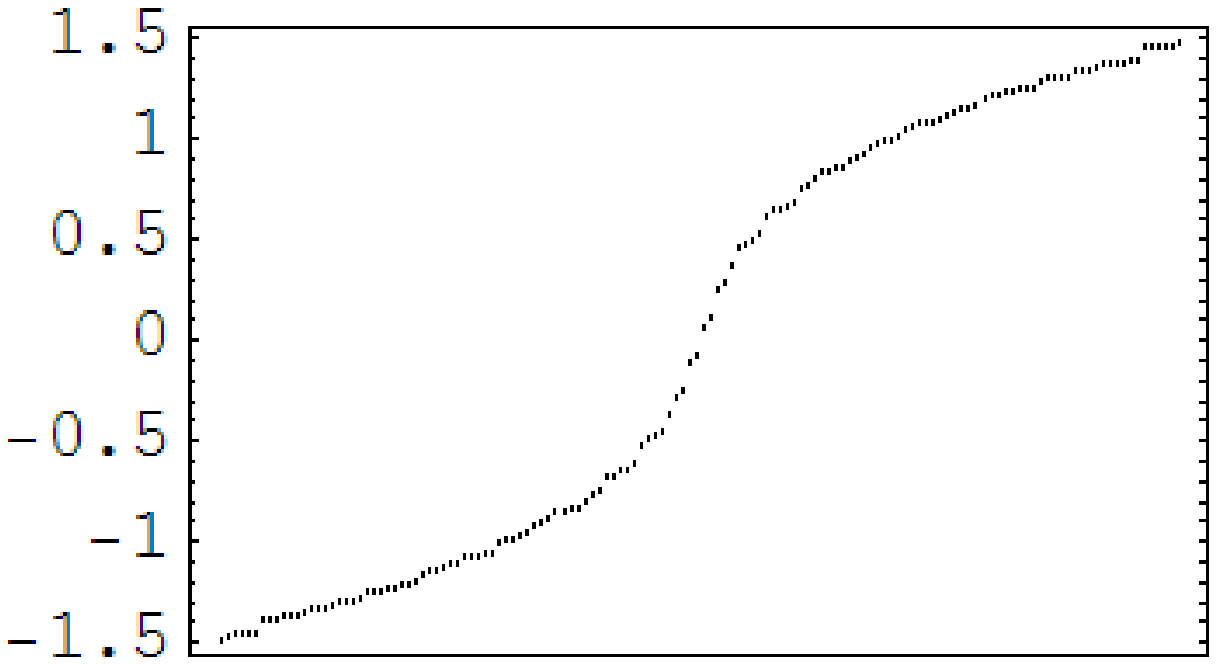}}
	\subfigure[]{\includegraphics[width=4cm]{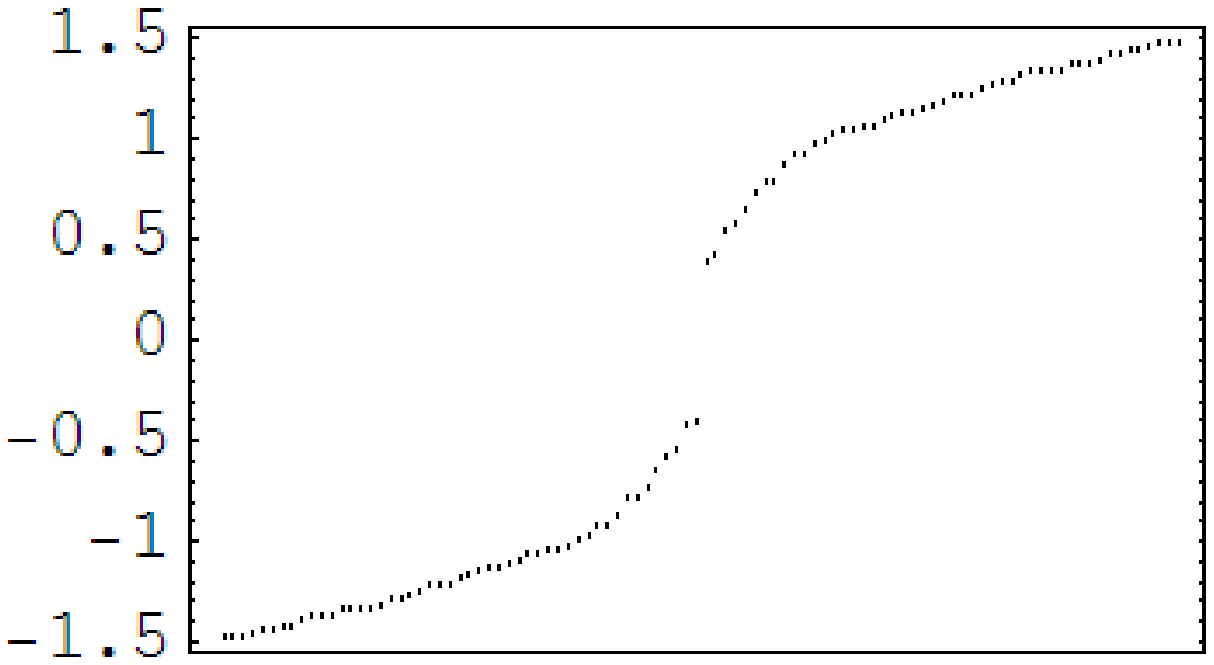}}
	\subfigure[]{\includegraphics[width=3.5cm]{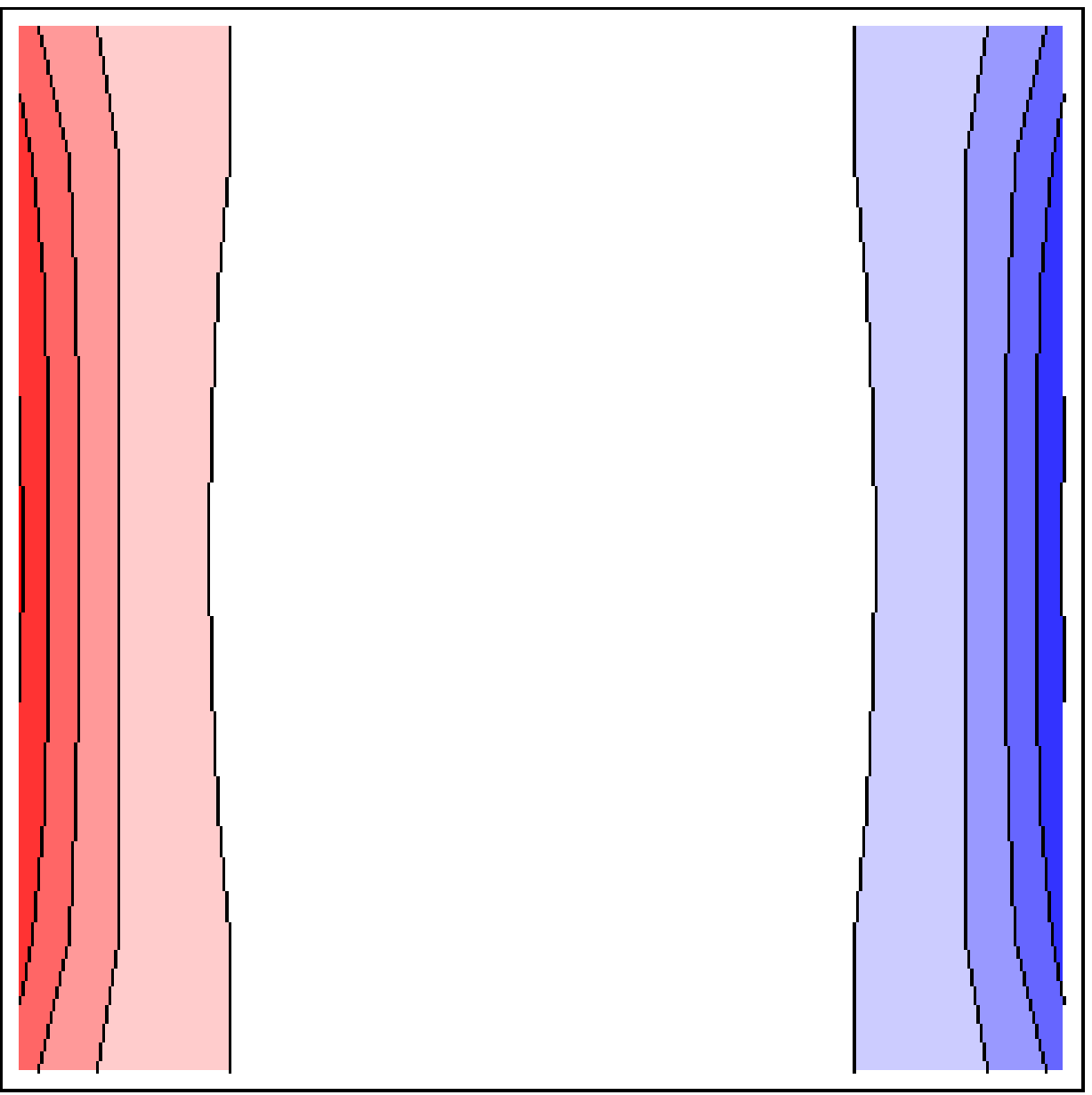}}
	\subfigure[]{\includegraphics[width=3.5cm]{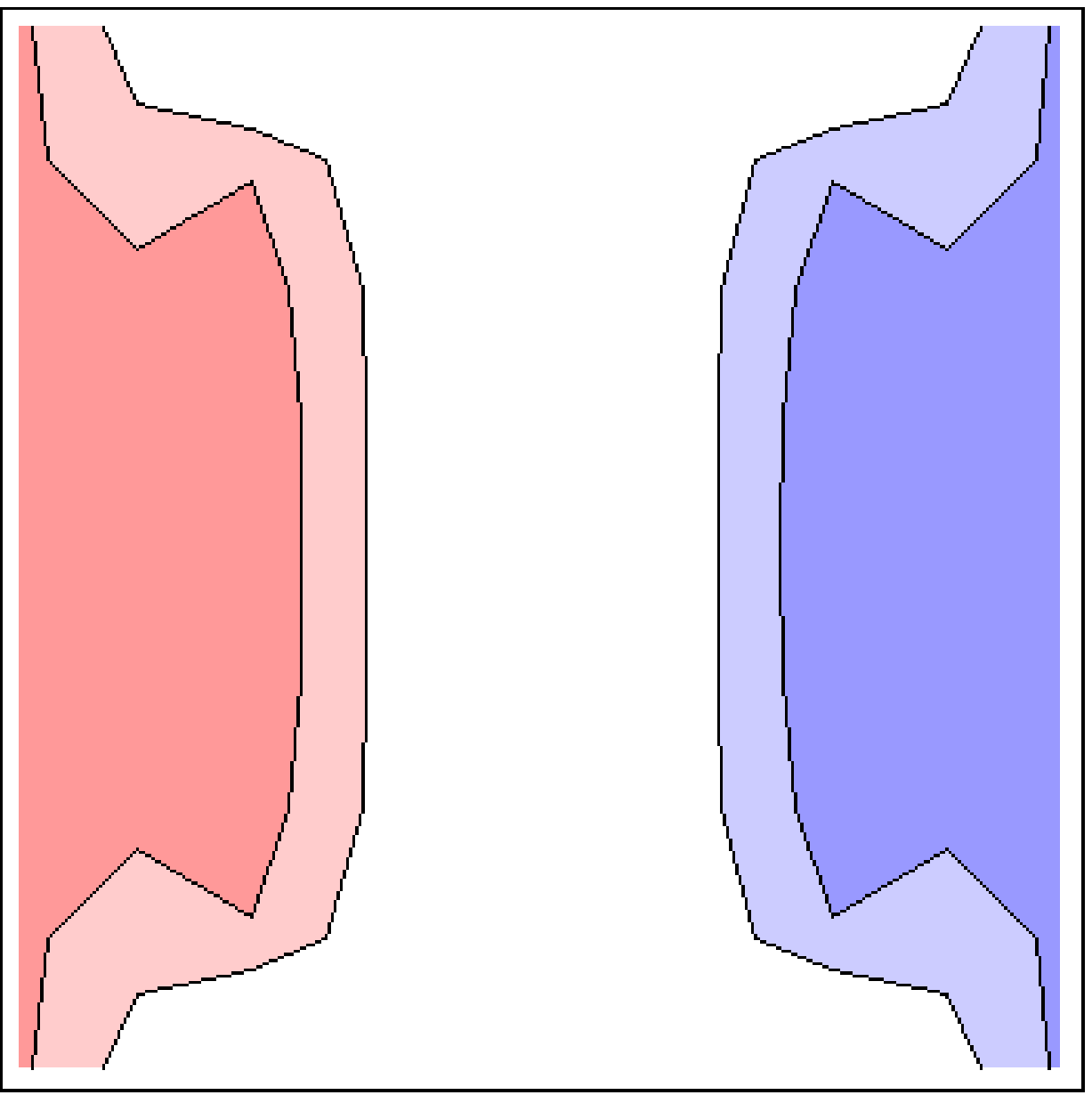}}
\subfigure{\includegraphics[width=0.9cm]{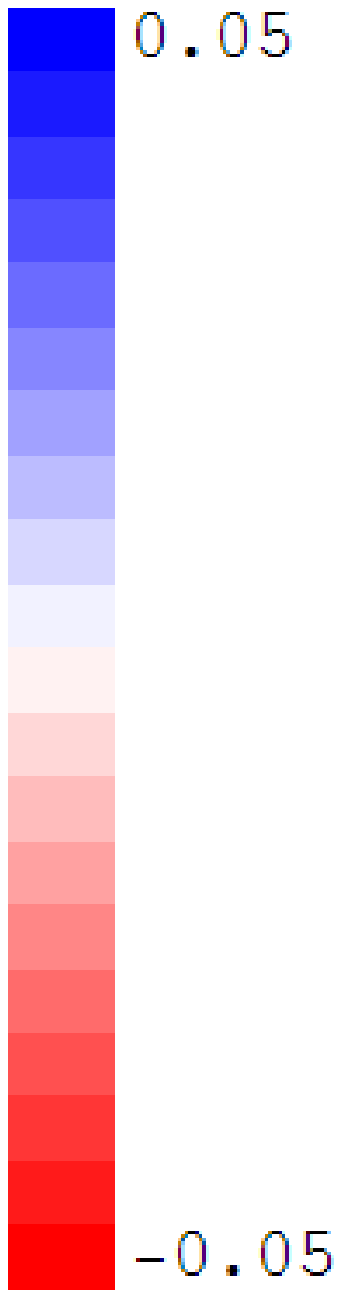}}	
	\caption{We now think of the previous model as from spin up electrons and pair it with its time reversal. We applied a unifrom Zeeman field $\delta H_z=0.2\sum_n c_n^\dag S_zc_n$. (a) We plot the eigenstate energies in ascending order. The edge states live inside the gap. (b) By applying a time-reversal as well as $S_z$ symmetry breaking term near the boundary $\delta H=\sum_{\rm n\in edge}c^\dag_n S_x c_n$, we can gap out the edge states. (c)-(d) We look at the current on the vertical links. While the current distributes slightly differently with or without the symmetry breaking term at the edges, the contributions to the bulk magnetization are identical.}
	\label{spinhall}
\end{figure}

The numerical result is shown in Fig. \ref{spinhall}. Here we can see that even though the edge states are gapped out by the local perturbations, the total current flowing near the edge remains the same. The local perturbation transfers the current from the states at the Fermi level, to the occupied bands. In the end, while local properties can affect the gapless states, the total current near the edge in the is unaffected.

We therefore conclude that the orbital magnetization, as well as the orbital-Zeeman susceptibility is independent of the boundary for an insulator. While the circulating current may be carried by the edge states, the total amount is entirely insensitive to the local boundary conditions. One can understand this from a calculation with periodic boundary conditions: the magnetization is calculated as an energy density in a magnetic field. The total energy, unlike the Berry's phase, is a truly extensive property, so that the boundary contribution is irrelevant in the thermodynamic limit. The energy density in the bulk is thus entirely independent of the boundaries far away enough, whether there are gapless states or not.


\section{magneto-electric effect} 
After the discussion of the polarization and the magnetization and seeing that they are thermodynamic quantities with very different behaviors, it is thus a natural question to ask the same question about the ME tensor; in addition, about how the Maxwell relation can be maintained. Before going into details of the boundary dependence, however, let us first show that the anisotropic part $\alpha_{3d}$ is independent of the boundaries.

In terms of electronic Green's functions and with periodic boundary conditions, we have derived the ME tensor from the OMP perspective, as a Berry's phase in a magnetic field:\cite{Tim2011c}
\bea\label{ompg}
\alpha_{ij}&=&(\alpha_{\rm wzw}+\alpha_{\rm 3d})_{ij},\nonumber\\
{\alpha_{\rm wzw}}_{ij}&=&-\frac{\pi i}6 \ep_{abcd}\Tr^S({\bf g}\p_a{\bf g}^{-1}{\bf g}\p_b{\bf g}^{-1}{\bf g}\p_c{\bf g}^{-1}{\bf g}\p_d{\bf g}^{-1}{\bf g})\delta_{ij};\nonumber\\
{\alpha_{\rm 3d}}_{ij}&=&-\frac{i}6\ep_{abj}\Tr\big(g\p_ig^{-1}g\p_ag^{-1}g\p_bg^{-1}g-h.c.\big).
\eea
The traces include the frequency and momentum integral divided by factors of $(2\pi)$; the symbol $\Tr^S$ denotes the integral and trace in one extra dimension in momentum space, with the original Brillouin zone and a trivial test system as the boundary. While the entire ME tensor is derived as a Berry's phase, $\alpha_{\rm 3d}$ does not depend on the Green's function extended to the extra dimension. Without considering boundaries directly, we can show that $\alpha_{\rm 3d}$ is independent of the boundaries, by showing it extends smoothly to finite frequency and momentum. 

At finite frequency and momentum, the ME response is understood as a term in the effective action which is proportional to $E^i(q,\omega)B^j(-q,-\omega)$. Unlike the uniform ME response however, this term can no longer be understood as OMP or OES, due to the fact that unlike uniform electromagnetic fields, the electric and magnetic fields at finite frequency and momentum are related by Faraday's law. The term nevertheless affects properties of the propagating electromagnetic waves. For our purposes, it suffices to show that the effective Lagrangian is continuous from $q=0$ to $q\rightarrow 0$. At any $q\neq 0$, we can calculate the effective Lagrangian by the conventional diagrammatic method. Calculated in the Appendix, the bubble diagram gives 
\beq
S_{ME}
=-\int\frac{{\rm d^4}q}{(2\pi)^4} B^\ell(q) E^k(-q)\alpha_{{\rm 3d}k\ell}+\mathcal{O}(q).
\eeq
Comparing with Eq.~(\ref{ompg}), we see that $\alpha_{\rm 3d}$ is continuous, whereas $\alpha_{\rm wzw}$ is entirely absent at finite momentum. One might worry that we have missed $\alpha_{\rm wzw}$ in momentum space due to the fact that it is a total derivative in real space, which Fourier transforms to zero and cannot be seen in momentum space. However, one can evaluate the diagram in real space, and it is still absent. Fundamentally this is due to the fact that the conventional perturbation theory is perturbative in orders of the gauge field, which breaks down with uniform field strength. Nevertheless, combining the two calculation, we can still say that $\alpha_{\rm 3d}$ is a bulk property and is independent of the boundaries. $\alpha_{\rm wzw}$, on the other hand, is similar to the polarization: it does depend on the boundary, but when there is no boundary, it presents itself as a Berry's phase. Note that one benefit of using the Green's function is that the separation of the local terms and boundary terms matches exactly how the expression depends on the extra dimension or not. This is not the case if we use the density matrices, either to calculate the same Berry's phase\cite{Tim2011a}, or to calculate a current response to a pumping procedure\cite{Essin2010}. In both calculations the ME tensor naturally separates into two terms, with the first term independent of the energy gap: 
\beq
\alpha=\alpha_{\rm cs}+\alpha_{\rm G};
\eeq
$\alpha_{\rm cs}$ is isotropic, but $\alpha_{\rm G}$ is not traceless. While $\alpha_{\rm G}$ can be uniquely determined by the bulk band structure and is independent of the boundaries, its trace is actually not measurable in the bulk.

Let us now focus at the isotropic part $\alpha_{\rm wzw}$. In terms of polarization in a magnetic field, the ambiguity is no surprise. However, how does the ambiguity of the orbital magnetization in a electric field come about?

One origin of the ambiguity is from the fact that the perturbation of a uniform electric field grows with distance. It therefore naturally depends on the boundary, when there is one. When we consider periodic boundary conditions, however, it becomes less clear.

In order to study the OES with periodic boundary conditions, we first have to properly define the magnetization with periodic boundary conditions. Without the current at the boundary, one sensible definition of the magnetization is from the relation $B=H+M$. That is, in the absence of applied current (which generates $H$), the magnetization simply equals the measured magnetic field. Note that with periodic boundary conditions and a finite volume, the magnetic field is quantized, because the total magnetic flux through the sample is quantized in units of $h/e$. In this case we take the perspective that the magnetic field will take the closest quantized value to the magnetization while the magnetization itself is still continuous. 



In our previous work\cite{Tim2011a}, we have shown that in a magnetic field, the $\theta$ term, which characterizes the isotropic part of the OMP, changes the quantization condition of the global electric flux. The ground state of the system thus carries an electric flux of -$(\theta e^2/2\pi h)\Phi_B+ne$, where $n$ is some integer that minimizes the flux. Using $0=D=E+P$, the $\theta$ term thus gives an isotropic orbital magneto-polarization response $\frac{\p P}{\p B}=\frac{\theta e^2}{2\pi h}$. However, this result is valid only when $(\Phi_B\theta e/2\pi h)<1$. In the thermodynamic limit this condition is always violated, and instead $\frac{\p P}{\p B}=0$. 

 Similarly, to see whether the same term contributes to the OES of the system, we would like to investigate whether there is a uniform magnetic field, when we constrain the path integral to have a given average electric field in the same direction. However, the electric field and the magnetic field behave in intrinsically different ways, when we formulated our theory assuming the existence of electric charges and the absence of magnetic monopoles: the quantization of the electric flux can change in the presence of the magnetic field, while the quantization of the magnetic flux is fixed at $(h/e)$. When we apply an electric flux, we can always imagine that the system is a coherent state composed of states with integer electric fluxes. The background magnetic field therefore does not have to be different from zero. Therefore, even at finite size, the $\theta$ term does not give rise to the OES. \textit{The Maxwell relation between the the isotropic OMP and the OES are thus violated.} They are only equal in the thermodynamic limit, where the $\theta$ term gives no contribution for both quantities. In other words, the isotropic OES is better thought of as a bulk-induced surface response, which vanishes when there is no boundary surfaces.

Now let us consider geometry with boundaries in some detail. From the result of Ref.~\cite{Malashevich2010}, we know that with open boundary conditions in all directions, the OES has an ambiguity only determined by specific surface boundary conditions. We have also seen in the introduction that in a cylinder geometry, the ambiguity of the OES can come from the quantized Hall current on the side surfaces. 

\begin{figure}[htb]
	\centering
	\includegraphics[width=8cm]{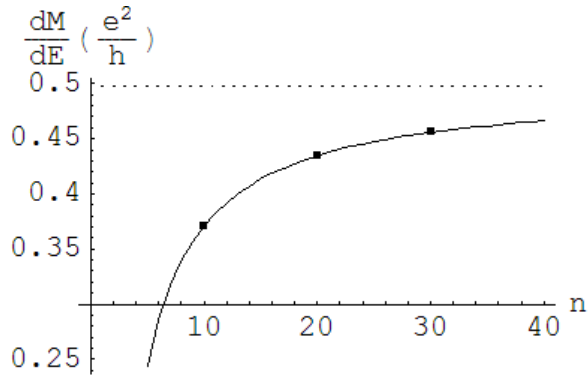}
	\caption{Here we plot the calculated OES versus the number of layers in the $z$-direction, with the model described by Eq.~(73) in Ref.~\cite{Qi08} with $\theta=0$, $m=c=1$. We take $\theta=0.5\pi$ at the top and the bottom layer to gap out the edge states. (If we take $\theta=\pm 0.5\pi$ on the two surfaces respectively, the whole system will be a Chern insulator and can no long be kept at charge neutrality without closing the gap in a magnetic field.) We put on an electric field such that the potential difference between the top and the bottom layer is $0.2$. The boxes show the calculated values. The solid curve is a fit by assuming a fixed width $w$ of the surface charges when there is a magnetic field, such that $\frac{{\rm d}M}{{\rm d}E}\propto(1-\frac wn)$. The fit gives gives $\frac{{\rm d}M}{{\rm d}E}=0.50 \frac{e^2}{h}$ in the thermodynamic limit and $w=2.54$. The OES changes sign as expected, when we change to $\theta=-0.5\pi$ instead on the boundary.}
	\label{OES}
\end{figure} 

What if there are no side surfaces? Suppose we take periodic boundary conditions only in two directions to get rid of the side surfaces. Does the OES still have the same ambiguity? One naively would expect the situation to be similar to the case with periodic boundary conditions, due to the absence of the possible circulating Hall currents. However, a more careful argument shows it is not the case. In fact, the system will spontaneously generate a magnetic field, which will then generate surface charge density $\sigma=\pm(\nu+\theta/2\pi)e^2B/h$ via the OMP response, to lower the electric energy. Minimizing the total energy as a function of $B$, we then get $B=M=(\nu+\theta/2\pi)e^2E/h$. While at finite size the total magnetic flux is quantized in units of $h/e$ in this setup, in the thermodynamic limit, the magnetic field will converge to the expected value, in contrast to the situations with periodic boundary conditions where it stays at zero. We have numerically confirmed this result by calculating the magnetization in the electric field, using the momentum space formula for the magnetization, derived in Ref.~\cite{Ceresoli06}, as shown in Fig.~\ref{OES}.

Before summing up, let us consider how gapless surface states can alter the ME response. Evidently, if we attach a fractional quantum Hall state on the side of the cylinder, the OES is going to change by a fraction of $e^2/h$.\cite{Brian2011} In general the fraction is quite arbitrary, so in this case the bulk value of the isotropic OES is not valid. This corresponds to the fact that the fractional quantum Hall state has ground state degeneracy. In general, we will therefore expect any gapless surface state will destroy the bulk description of the isotropic ME response.

To sum up, The anisotropic part of the ME tensor $\alpha_{\rm 3d}$ is independent of the boundaries. The isotropic part $\alpha_{wzw}$ depends partially on the boundary. While $\alpha_{\rm 3d}$ is a truly local quantity, $\alpha_{\rm wzw}$ only lives at $q=0$. Corroborating with the fact that both isotropic OES and OMP responses vanish with periodic boundary conditions in the thermodynamic limit, it is better to  think of $\alpha_{\rm wzw}$ as a quantized surface effect induced by the bulk.   

\section{conclusion}
 We have thus gone through polarization, magnetization, and magneto-electric responses and see their dependence on the boundary. 
A bulk calculation done with periodic boundary conditions contains enough information to predict how the quantity in question can depend on the boundary, however. In particular, using our formalism described in Ref.~\cite{Tim2011c}, any quantity that does not involve an extension of the Green's function to one extra dimension is independent of the boundary. On the other hand, quantities that requires an extension to extra dimension will depend on the boundary. The bulk can determine its value up to some quantized amount, only when (i) there are no gapless surface states, (ii) surfaces break no symmetry that is required to determine the bulk value with periodic boundary conditions, and (iii) the system is kept at charge neutrality. If any of the conditions are violated, the surface contribution will dominate and render the results obtained with periodic boundary conditions invalid.

\section*{Acknowledgement}
We thank X.G. Wen and N. Nagaosa for insightful discussions. This work is supported by NSF grant DMR 1104498.
\bibliography{mybib}

$\;$
\newpage 
\newpage

\begin{widetext}
\section*{Appendix: ME Effect at Finite Momentum}

In the main text we have argued heuristically that the trace of the ME tensor comes entirely from the surface, and therefore does not contribute at finite momentum. The locally measurable ME tensor is therefore traceless in the $q\rightarrow 0$ limit. We can directly calculate the ME tensor at finite $q$: Fourier transforming and expanding the hopping Hamitonian up to second order of $A^\mu$, we have
\beq
\label{H}
\Delta H=\sum_{k,q}c^\dag_{k+q/2}\p_uH(k)c_{k-q/2}A^\mu(-q)+\frac12\sum_{k,q,q'}c^
\dag_{k+(q+q')/2}\p_u\p_\nu H(k)c_{k-(q+q')/2}A^\mu(-q)A^\nu(-q');
\eeq
$H_k\equiv\sum_{d_i}t_{d_i}\exp(ikd_i)$ is a matrix. Integrating out the electrons, the effective action at quadratic order of $A_\mu$ reads

\beq
S_{\rm eff}=\int\frac{{\rm d^4}q}{(2\pi)^4}\frac i2 A^\mu(q) A^\nu(-q)\Tr\bigg(\p_\mu\p_\nu g^{-1}(k)g(k)+\p_\mu g^{-1}(k)g(k+\frac{q}2)\p_{\nu}g^{-1}(k)g(k-\frac{q}2)\bigg);
\eeq

similarly, the trace includes the integral of energy and momentum divided by $2\pi$. The first term in the trace is from the second term in Eq.~(\ref{H}), usually called the paramagnetic current, and does not have $q$ dependence. To compare with Eq.~(\ref{ompg}), we Taylor-expand the second term to second order in $q$ to get the behavior in the $q\rightarrow 0$ limit: (From here on, we drop the dependence on $k$ to avoid cluttering.)
\beq
S_{\rm eff}\sim\int\frac{{\rm d^4}q}{(2\pi)^4}-\frac i4 q^\lambda q^\sigma A^\mu(q) A^\nu(-q)\Tr\bigg(\p_\mu g^{-1}\p_\lambda g\p_{\nu}g^{-1}\p_\sigma g-\frac12\p_\mu g^{-1}\p_\lambda\p_\sigma g\p_{\nu}g^{-1} g-\frac12\p_\mu g^{-1}g\p_{\nu}g^{-1}\p_\lambda\p_\sigma g\bigg)+\mathcal{O}(q^3);
\eeq

To further simplify the expression, let us now take the Coulomb gauge. In the Coulomb gauge, we have to take either $\lambda$ or $\sigma$ to be in the time direction to have the expression contribute to the ME tensor. Since $\p_i\p_\omega g^{-1}=0$, we can integrate-by-part the time derivative. using $\p_\omega g=-g^2$, and rename the indices $i, j, k$, now running through only the spatial directions, we get
\beq\label{sme}
S_{ME}\sim\int\frac{{\rm d^4}q}{(2\pi)^4}\frac i2 \omega q^i A^j(q) A^k(-q)\Tr\bigg(g\p_j g^{-1}\p_i g\p_{k}g^{-1} g\bigg).
\eeq
Now we need to massage the expression a little bit. Let us use $(ijk)$ as a short-hand notation of the expression $\Tr\bigg(g\p_i g^{-1}\p_j g\p_{k}g^{-1} g\bigg)$. Integrating by parts\cite{Tim2011c}, we have the following relation:
\beq
(ijk)+(jki)+(kij)=0.
\eeq
We therefore have
\beq
(jik)=\frac23\big(2(jik)-(ikj)-(kji)\big).
\eeq
In the trace in Eq.~(\ref{sme}), only the part symmetric under the exchange of the index $j$ and $k$ would contribute, as we can change variables from $q$ to $-q$, effectively exchanging $A^j(q)$ and $A^k(-q)$. Therefore, in the expression above, we can exchange $j$ and $k$ freely. We therefore have
\bea
S_{ME}&\sim&\int\frac{{\rm d^4}q}{(2\pi)^4}\frac i{6} \omega q^i A^j(q) A^k(-q)
\big((jik)+(kij)-(ijk)-(kji)\big)\nonumber\\
&=&\int\frac{{\rm d^4}q}{(2\pi)^4}\frac i{6}\omega q^i A^j(q) A^k(-q)\ep_{ij\ell}\ep_{ab\ell}\big((kab)+(bak)\big).\nonumber\\
&=&\int\frac{{\rm d^4}q}{(2\pi)^4}\frac i{6} B^\ell(q) E^k(-q)\ep_{ab\ell}\big((kab)+(bak)\big)\nonumber\\
&\equiv&-\int\frac{{\rm d^4}q}{(2\pi)^4} B^\ell(q) E^k(-q)\alpha_{k\ell}(q\rightarrow 0).
\eea

$\alpha_{k\ell}(q\rightarrow 0)=-\frac i{6}\ep_{ab\ell}\big((kab)+(bak)\big)$ is traceless, as the two terms cancel each other with antisymmetrization. 
\end{widetext}
\end{document}